\documentclass[5p]{elsarticle}

\usepackage{lineno,hyperref}
\usepackage{dblfloatfix} 
\usepackage{float}
\usepackage{graphicx}
\usepackage{soul} 
\usepackage{xcolor} 
\usepackage{caption}

\usepackage[detect-all]{siunitx}
\newcommand{\mcr}{\si{\micro}} 

\journal{arXiv.org}

\begin{document}


\begin{frontmatter}

\title{Anomalous High Strain Rate Compressive Behavior of Additively Manufactured Copper Micropillars}

\author[mymainaddress,address3]{Rajaprakash Ramachandramoorthy \corref{mycorrespondingauthor1}}\cortext[mycorrespondingauthor1]{Corresponding author, equal contribution} \ead{r.ram@mpie.de}

\author[mymainaddress,myaddress2]{Szilvia Kal\'{a}cska \corref{mycorrespondingauthor2}}\cortext[mycorrespondingauthor2]{Corresponding author, equal contribution} \ead{szilvia.kalacska@cnrs.fr}

\author[address4]{Gabriel Poras}
\author[mymainaddress]{Jakob Schwiedrzik}
\author[mymainaddress]{Thomas E. J. Edwards}
\author[mymainaddress]{Xavier Maeder}
\author[address5]{Thibaut Merle}
\author[address5]{Giorgio Ercolano}
\author[address5]{Wabe W. Koelmans}
\author[mymainaddress]{Johann Michler}

\address[mymainaddress]{Empa, Swiss Federal Laboratories for Materials Science and Technology, Laboratory of Mechanics of Materials and Nanostructures, CH-3602 Thun, Feuerwerkerstrasse 39. Switzerland}
\address[address3]{Current Address: Max-Planck-Institute für Eisenforschung, Max Planck Strasse 1, 40472 Düsseldorf, Germany}
\address[myaddress2]{Current Address: Mines Saint-Etienne, Univ Lyon, CNRS, UMR 5307 LGF, Centre SMS, 158 cours Fauriel 42023 Saint-Étienne, France}
\address[address4]{ESPCI Paris, 10 rue Vauquelin, 75005 Paris, France}
\address[address5]{Exaddon AG, Sägereistrasse 25, 8152 Glattbrugg, Switzerland}

\begin{abstract}
Microscale dynamic testing is vital to the understanding of material behavior at application relevant strain rates. However, despite two decades of intense micromechanics research, the testing of microscale metals has been largely limited to quasi-static strain rates. Here we report the dynamic compression testing of pristine 3D printed copper micropillars at strain rates from 
$\sim0.001$ s$^{-1}$ to $\sim500$ s$^{-1}$. It was identified that microcrystalline copper micropillars deform in a single-shear like manner exhibiting a weak strain rate dependence at all strain rates. Ultrafine grained (UFG) copper micropillars, however, deform homogenously via barreling and show strong rate-dependence and small activation volumes at strain rates up to $\sim0.1$ s$^{-1}$, suggesting dislocation nucleation as the deformation mechanism. At higher strain rates, yield stress saturates remarkably, resulting in a decrease of strain rate sensitivity by two orders of magnitude and a four-fold increase in activation volume, implying a transition in deformation mechanism to collective dislocation nucleation. 

\end{abstract}


\end{frontmatter}


\section{Introduction}

The recent push towards novel miniaturized applications (\emph{e.g.}, small scale sensors, actuators, printed circuit elements \emph{etc.}) has resulted in the pursuit of new technologies capable of manufacturing damage-free, device-grade microarchitectures. For example, a recent combination of laser exposure and chemical etching was used to produce glass microparts and two photon lithography was applied for polymer microprinting with nanoscale resolution \cite{Lenssen.2012, Meza.2015}. In case of metals, unfortunately, the typical additive manufacturing-based 3D printing methods such as selective laser sintering, direct/laser metal deposition and metal powder bed fusion lack the resolution to fabricate micron-to-meso scale complex metal samples \cite{SLM, Karunakaran.2012, BEAM, Beiker.2014}. A common problem with 3D printing in general is that due to the sequential printing of components each sample has potentially a different microstructure and distribution of manufacturing flaws. This requires usually a statistically significant number of samples to be analyzed to gain an in-depth understanding of the process. Furthermore, most micromechanical testing studies rely on samples prepared using laser, ion beam, reactive ion etching \emph{etc}. leading to a damaged surface layer that influences the deformation behavior \cite{Kiener.2007, Gigax.2019}. Recent reviews highlight the few techniques that are available for manufacturing small-scale metal samples, including traditional FIB milling, template-based electrodeposition and a few recent additive micromanufacturing techniques \cite{Daryadel.2019, Reiser.2020}. 

The template assisted electrodeposition (TAE) fabrication method is a two-step process that involves fabricating a polymer/anodic aluminium oxide (AAO) negative mold with holes of the microarchitectures and subsequent electroplating into small-scale orifices and stripping of the mold. Recently, Schürch \emph{et al.} showed that this method could be successfully used on arrays of nanocrystalline micropillars and microsprings \cite{Schurch.2020}. Though it is a unique method of metal microarchitecture manufacturing, it is a multi-step process of high complexity that severely limits the design freedom \cite{Greer.2015}.

Additive micromanufacturing (\mcr AM) typically involves delivery of metals using a hollow microcantilever (lithography made silicon or stretched glass micropipette), with a shape similar to an atomic force microscopy (AFM) tip, with a nanoscale aperture. Further, there are two broad classifications in the techniques available for the out-of-plane growth of these metal microvoxels, namely the i) two-step colloidal ink direct writing method that involves a thermal annealing step after writing and ii) electrochemical methods that use electrochemical reduction of metal ions for direct metal writing.  Electrochemical methods, such as meniscus confined electrodeposition, in-liquid electrodeposition and electrohydrodynamic redox printing, are preferred, as they provide dense and crystalline metal microarchitectures directly, without any post-print processing. There are a few recent studies on using these technologies to create simple, metal microarchitectures such as micropillars with sample dimensions ranging from $\sim$1 \mcr m to $\sim$10 \mcr m \cite{Gu.2012, Moestopo.2020}. Specifically, Daryadel \emph{et al.} have used a meniscus-confined pulsed electrodeposition method to create nanotwinned copper micropillars of $\sim$700 nm diameter \cite{Daryadel.2018, Daryadel.2020}. Similarly, Raiser \emph{et al.} have reported the fabrication of sub-micron scale ($\sim$400 nm diameter) multi-metal pillars using electrohydrodynamic redox printing \cite{Reiser.2019}. In addition, it should be noted that several of these studies report metal structures with rough sidewalls and non-flat tops, which are non-ideal for mechanical characterization. 

A core reason to develop \mcr AM is the envisioned use of the fabricated microscale parts in real-life applications. In such applications, the microscale parts are expected to withstand a variety of stress-strains, including vibrations, drops, thermal expansions, external impacts, and penetrations. These requirements demand an extensive exploration of rate-dependent, micromechanical properties from quasi-static strain rates, up to $\sim$1000 s$^{-1}$ (to simulate impacts) on these small parts \cite{Jennings.2011, Raj.2019}. As such, mechanical properties of metals with different microstructures and defect structures such as ultrafine grain, coarse grain, nanotwinned and nanocrystalline have been of considerable interest over the last few decades \cite{Mieszala.2016}. Surely, in the macroscale mechanics community there is a plethora of information available on the quasi-static and high strain rate properties of microstructure-dependent mechanical properties of metals \cite{Jia.2001}. Meanwhile, the micromechanical properties of metals (characteristic dimension $<$10 \mcr m) that are typically ascertained using micropillar compression or microtensile tests have been limited to quasi-static strain rates \cite{Guillonneau.2018}. Therefore, the experimental quantification of the mechanical properties of metals at the microscale remain largely unexplored at strain rates above 0.1 s$^{-1}$ \cite{Jennings.2011, Raj.2016, Raj.2018, Li.2020}. Recently, the authors of this study reported the micromechanical properties of fused silica, bulk metallic glass and polymer micropillars at strain rates up to 1000 s$^{-1}$ using a piezo-based testing platform \cite{Raj.2019, Raj.2021}. There are also a small number of other recent key studies on epoxy\cite{Rueda.2020}, single crystal copper \cite{Breumier.2020} and magnesium micropillars \cite{Lin.2021} conducted at strain rates up to 100 s$^{-1}$ and nanoindentation of coarse-grained aluminium and nanocrystalline nickel performed at indentation strain rates up to 100 s$^{-1}$ \cite{Merle.2020}. 

The mechanical properties of copper with a variety of microstructure (microcrystalline (MC), ultra-fine grained (UFG), nanocrystalline and nanotwinned) have been studied extensively in the literature \cite{Jennings.2011, Raj.2019} at both macro- and microscale \cite{Wimmer.2014, Kim.2011, Valiev.1994}. In this study, we specifically focus on the mechanical properties of UFG and MC copper. Several macroscale mechanical studies have reported both the quasi-static and dynamic properties of UFG and MC copper across several orders of strain rate magnitude from 0.0001 s$^{-1}$ to 3000 s$^{-1}$ \cite{Wimmer.2014}. In general, all these macroscopic studies report a low strain rate sensitivity upto a strain rate of ~1000 s$^{-1}$ and a sudden increase of strain rate sensitivity at even higher strain rates owing to dislocation drag in copper \cite{Jia.2001, Andrade.1994}. But a critical point to note in these studies is that the experiments are typically conducted only between strain rates of 0.001 s$^{-1}$ till 0.1 s$^{-1}$ (through the application of universal testing systems) and at strain rates of 1000 s$^{-1}$ and higher (using Kolsky bars and split-Hopkinson pressure bars). Meaning there is usually a gap in strain rates between 0.1 s$^{-1}$ and 1000s$^{-1}$ where no experimental data is available \cite{Jia.2001, Gray.1997}. This is owing to the lack of instrumentation available to explore these strain rates, though a few servo-hydraulic systems have been designed to address this issue \cite{Oosterkamp.2000, Huh.2009}. As such, the mechanical properties of UFG and MC copper have not been established conclusively in this intermediate strain rate regime. In addition, due to the lack of consistency in microstructure/density across large macroscale samples, sensitive transitions in strain rate sensitivity can be missed \cite{Raj.2021, Xiao.2021}. To this effect, we chose micropillar-based compression testing with consistent local microstructure to investigate the mechanical properties of UFG and MC copper across a wide range of strain rates. 

We report here the additive micromanufacturing of copper micropillars and a thorough investigation of their microstructural and mechanical properties as a function of application-relevant strain rates. Samples were fabricated with a \mcr AM technique that uses localized electroplating and \emph{in situ} voxel completion detection \cite{Ercolano.2020}. The copper micropillars were fabricated with a cylindrical dogbone geometry with flat tops, making them ideal test beds for mechanical characterization \cite{Frenzel.2016}. Subsequently, using a recently developed piezo-based micromechanical testing platform (Alemnis AG), the rate- and microstructure- dependent compressive properties of these copper micropillars were identified across five orders of magnitude in strain rate from $\sim$0.001 s$^{-1}$ to $\sim$500 s$^{-1}$. 


\section{Material and Methods}
\subsection{3D printing of copper micropillars}

\begin{figure*}
  \includegraphics[width=0.99\linewidth]{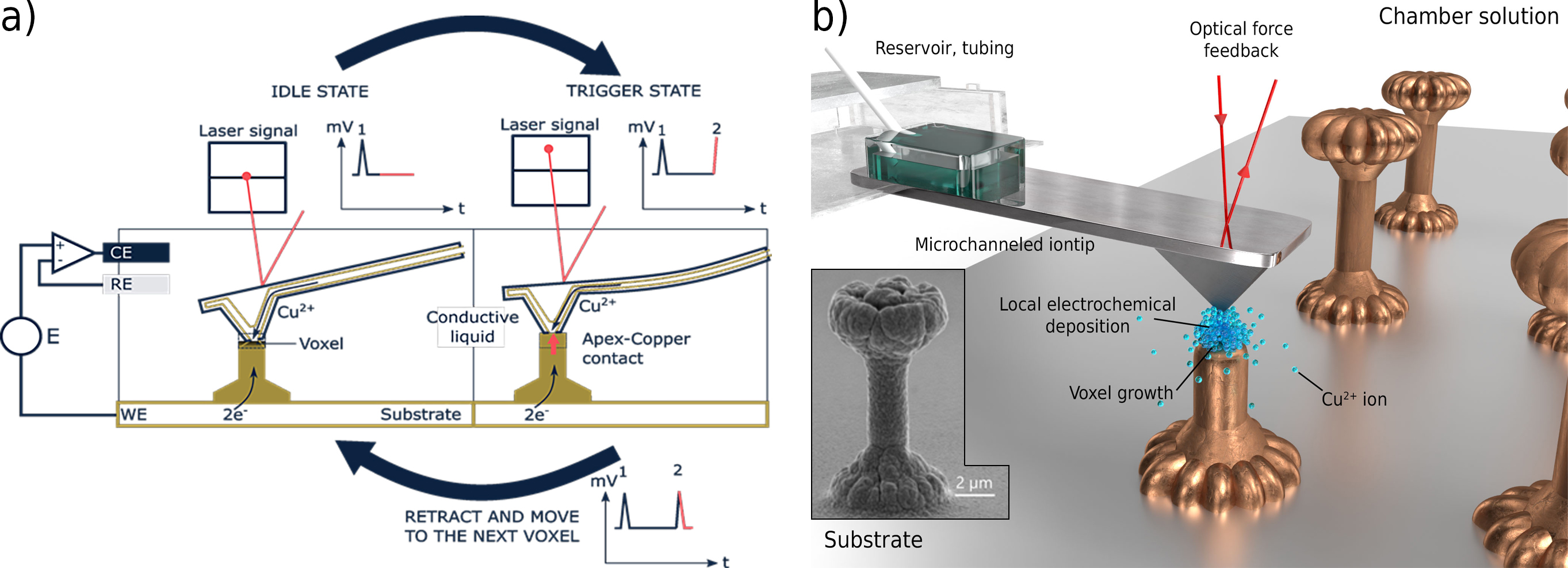}
  \caption{a) \mcr AM of metal micropillars using the localized and force-controlled electroplating technique in three-electrode cell configuration. A copper voxel grows underneath a suspended tip that expels copper ions. Voxel completion leads to a tip-voxel interaction that is registered by an optical beam detection system. (Abbreviations: WE - working electrode; RE - reference electrode; CE - counter electrode). b) 3D sketch of the printing process. Inset shows the resulting micropillar.}
  \label{fig:01}
\end{figure*}

Metal microarchitectures reported in this study were made by \mcr AM via a localized force-controlled electroplating technique using the CERES system (Exaddon AG, Switzerland). As shown by the schematics in Figure \ref{fig:01}, a hollow silicon nitride AFM tip is immersed in a standard three-cell electrochemical cell and the precursor copper ions are fed through the tip. The local ion supply confines the electrochemical reduction to the desired surface area of the working electrode. Initially, the tip is kept at a chosen distance (termed the voxel height) from the substrate, locally plating the metal. The metal deposition proceeds, until the metal deposit reaches the tip. The deposit interacts with the tip and slightly bends the cantilever up. An optical beam deflection system measures this deflection. As soon as a preset deflection threshold is reached (typically in the order of 1 nm), the voxel is considered complete, and the tip is moved to the next voxel coordinate. There is no post-processing done and the process is performed at room temperature. A detailed description of \mcr AM and its advantages over other techniques is given elsewhere \cite{Ercolano.2020b}. The air pressure that is used to expel the copper ion solution from the tip is variable and was used to control the lateral voxel size \cite{Frenzel.2016}. For the dogbone micropillars in this study, a tip with a 300-nm-diameter opening (Iontip, Exaddon AG, Switzerland) was used at a pressure range of 50 mbar to 200 mbar. Each dogbone is built up out of 104 voxels and the mean printing time of a pillar was $61 \pm 1$ s over the whole $10 \times 10$ array.

The copper micropillar array was fabricated with a spacing between the micropillars ideal for mechanical testing. In addition, the dogbone shape of the micropillars was chosen to avoid errors in the mechanical tests due to the rounding of the micropillars in the top, as shown in previous literature \cite{Reiser.2020}. The levelled top makes the pillars ideally suitable for compression testing with a diamond flat-punch \cite{Schurch.2020}. Supplementary Figure S1 clearly show the advantages of using \mcr AM, such as the well-defined shape and dimensions of the micropillars along with their reproducibility over a 100-pillar array. In addition, elemental analysis was conducted using energy-dispersive X-ray spectroscopy (EDS) mapping on a copper micropillar. The analysis confirms that the pillars have no additives/impurities and have no inhomogeneity in the distribution of chemical elements, like boundary segregation of elements (Supplementary Figure S2). 

The chemical composition used for this study consists of two solutions. The first is the conductive liquid that is present in the three-electrode cell as shown in Figure \ref{fig:01}. The conductive liquid is a 54 mM solution of sulfuric acid with 0.5 mM hydrochloric acid. The bath volume is about 40 ml. The printing electrolyte is 0.5 M copper sulphate solution in a 51 mM sulfuric acid solution with 0.48 mM of hydrochloric acid. The printing electrolyte is inserted into a small reservoir present in the consumable holder of the tip. The total volume of electrolyte in the reservoir is 1 \mcr l. The flow of electrolyte through the tip during printing is estimated to be $2 - 20$ fl s$^{-1}$.

\subsection{Micromechanical testing}

The micropillar compression was conducted using an \emph{in situ} micromechanical testing system (Alemnis AG, Switzerland) inside a Zeiss DSM 962 SEM. The system uses a piezostack-based actuator that can be moved with a maximum speed of $\sim$ 8 mm$\cdot$s$^{-1}$. For the quasi-static studies (strain rate $\leq$ 0.1 s$^{-1}$), a typical strain gage-based load cell was used to capture the forces. At higher strain rates, a piezoelectric load cell with a much higher stiffness ($k_{\textrm{HSR}}\sim 2.5 \times 10^{7}$ N$\cdot$m$^{-1}$) was used instead of the strain gage-based load cell ($k_{\textrm{QS}}\sim 2 \times 10^{5}$ N$\cdot$m$^{-1}$) to avoid ringing artifacts during high-speed movements. High fidelity load and displacement signals were captured using a support hardware, capable of a sampling rate of 1 MHz. A more detailed description of the testing system is mentioned in Ref. \cite{Raj.2021}.

\subsection{EBSD-based microstructural analysis of copper micropillars}

\begin{figure*}[!ht]
  \includegraphics[width=\linewidth]{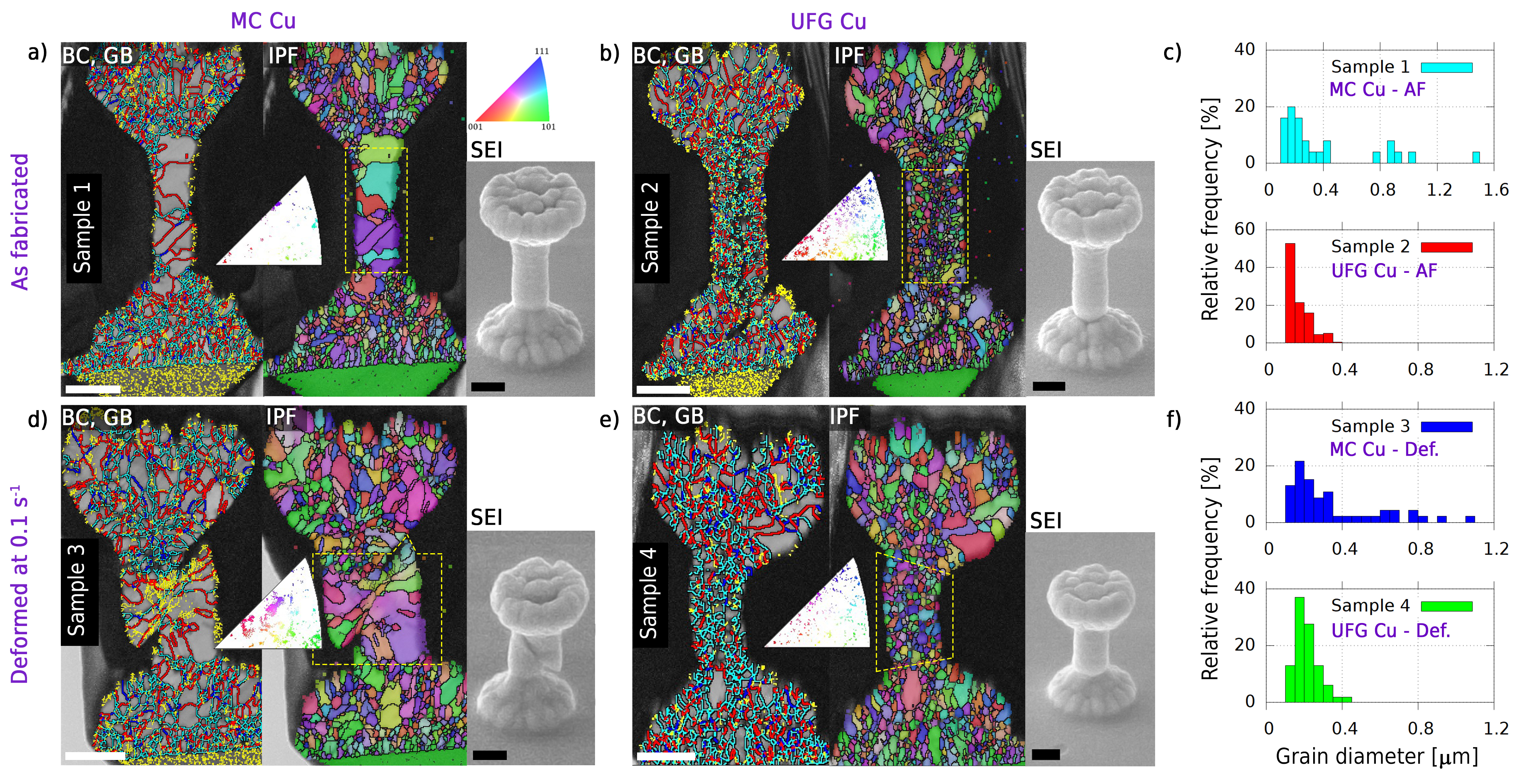}
  \caption{EBSD results measured on undeformed pillars with a) MC and b) UFG grain sizes and the corresponding deformed cases at a strain rate of 0.1 s$^{-1}$ shown in d) and e) with similar microstructure. Grayscale plots present band contrast (BC) values highlighting the grain structure, while colored lines mark the detected grain boundaries (GB); red: $\Sigma$3 twin, blue: $\Sigma$9 twin, azure: HAGB ($>$ 10$^{\circ}$), yellow: LAGB (1$^{\circ}$ -- 10$^{\circ}$). Yellow rectangle surrounds the gage section used for statistical analysis. Inverse pole figure (IPF) color maps show the corresponding orientation, while the inset IPF triangles plot the orientation of the grains only in the gage section. c) and f) plot the relative frequencies (provided in terms of percentage of total grain number within a specified size interval) of detected grain diameters in the gage sections of the pillars. The minimum area was set to 10 pixels and the grain distinction threshold was defined as 10$^{\circ}$ misorientation. The histogram class width is 0.05 \mcr m. Secondary electron images (SEI) show topological features on the surface of the pillars. Scale bar corresponds to 2 \mcr m.}
  \label{fig:02}
\end{figure*}

To investigate the initial microstructure prior to deformation, two samples were isolated from the Si substrate by the lift-out process (Figure \ref{fig:02} a, b). For the process, a Tescan Vela Ga$^+$ focused ion beam (FIB) SEM was used. The SEM was equipped with an Orsay Physics Pt gas injection system. The full process is shown in Supplementary Figure S14. First, a platinum cap was deposited on top of each pillar to reduce damage from consecutive FIB imaging. Then a trench around the pillar was milled, while the substrate material redeposited around the sample. This redeposition created a second protective layer that conserved well the inner structure for further measurements. After the needle of an Omniprobe nanomanipulator was attached to the sample, the final FIB cut was made, and the pillars were attached to the side of a TEM sample holder by Pt deposition.

Afterwards, the samples were measured in a Tescan Lyra3 FIB-SEM that was equipped with an Oxford Instruments Symmetry EBSD camera. In this system the vacuum chamber is specially designed to allow FIB milling and EBSD measurement without moving the sample \cite{Kalacska.2020}. This static setup allowed us to carefully monitor the progress of the FIB cross sectioning with a 30 kV, 0.6 nA rough beam and prepare multiple parallel FIB cuts to study several cross sections of the same sample. Before the final cut, the current was reduced to 65 pA and the subsequent EBSD measurement was performed with a 20 kV, 7 nA electron beam. The mapping step size was $30 - 40$ nm, while diffraction patterns were recorded with a $2 \times 2$ binning. Data analysis was performed with HKL Tango and Mambo (v.5.12).

HR-EBSD evaluation was conducted on the deformed samples to estimate stress localization along the pillars. HR-EBSD utilizes a cross-correlation-based image analysis on the collected diffraction patterns \cite{Britton.2011}. For $\sigma_{33}$ a traction-free boundary condition was assumed \cite{Wilkinson.2006}. This simplification is justified by the small information volume from which diffraction patterns originate. The depth of the surface from where the electrons scatter and reach the EBSD detector is in the order of 20 - 50 nm\cite{Chen.2011}, therefore the stress tensor components directed outwards of the surface plane must equal to zero. In each grain a reference pattern was chosen based on the lowest kernel average misorientation (KAM) value, which is associated with a lower stress level. Each pixel within the grains was related to these reference patterns, therefore the scales are relative. For the HR-EBSD analysis, BLG Vantage CrossCourt v4 was used. As this technique requires a high-quality reference diffraction pattern from each grain, areas with low image quality were removed from the evaluation. Because of this, only the sample with the biggest grains was studied this way. Note that if a reference pattern with the lowest KAM value is not related to the minimal stress state, all stress values in that grain will appear abnormally high (as demonstrated in Figure \ref{fig:04} on one grain in the top left corner).

\subsection{EDS-based chemical analysis of copper micropillars}

EDS characterization was performed on Sample 1 (Figure \ref{fig:02}a) using an Oxford Instruments Ultim Max 65 detector with a 20 kV electron beam. The identified chemical maps are shown in Supplementary Figure S2. The micropillar itself has a homogeneous Cu distribution, only a slight curtaining effect from FIB milling is visible in the Cu K$\alpha$ map. Due to the lift-out procedure, Ga, Pt and Si from the substrate was deposited around the pure copper pillar. C is only present on the surface of the sample, which possibly originated from the precursor gas for Pt deposition prior to FIB milling, as its distribution is consistent with the Pt M$\alpha$ map. The presence of O was estimated to be less than 0.2 wt\% inside the pillar, which is close to the detection limit for this technique. The presence of Cl or S was not detected in the manufactured samples, that could have originated from the bath. It is noted that EDS at this scale is not suitable to address boundary segregation issues, and the technique was only applied to get a larger scale overview and to support the homogeneous nature of the samples.

\subsection{TKD and TEM-based microstructure characterization}
TKD and TEM measurements were taken on deformed micropillars, after typical lamella thinning was carried out using sequential FIB polishing starting from 30 kV 0.9 nA decreasing to the final 5 kV 70 pA milling. TKD analysis was performed using 30 kV 16 nA electron beam in a Tescan Lyra3 SEM. Higher resolution imaging was performed on a transmission electron microscope (JEM 2200fs, JEOL, Japan) at 200 kV operated in both projection (TEM) and scanning probe (STEM) modes.

\subsection{FEM-based analysis}

A FEM-based thermal analysis was conducted by COMSOL Multiphysics software 5.5 to understand the extent of adiabatic heating in the micropillars based on the power input during a high strain rate test. The modelling was set up on a circular Cu pillar on a Si substrate. The plastic dissipation was modelled as volumetric heat source in the pillar's gage section (257756.6 MJ$\cdot $m$^{-3}\cdot s^{-1}$)\cite{Kapoor.1998} with a duration of 1 ms (similar to the experiment). Boundaries in the half space below were modelled as constant temperature boundaries, where the radius was 5 times larger than the pillar's base radius. Free boundaries were modelled as insulating boundaries with no radiation. The top boundary (in contact with the indenter) was modelled as either insulating boundary (upper bound) or constant temperature boundary (lover bound). Further discussion on the simulation can be found in Supplementary Section SS6.

In order to see the effect of the dogbone shape on the elastic stress distribution and deformation behaviour of the micropillars, an axisymmetric simulation was prepared using the actual dimensions of the samples. The elastic moduli were set to $E_{Cu}=110$ GPa, $E_{Si}=170$ GPa, and Poisson's ratios of $\nu_{Cu}=0.35$ and $\nu_{Si}=0.28$ were selected from COMSOL's standard materials' library. The geometry was meshed with 11223 domain elements and 512 boundary elements. Boundaries were fixed at the bottom and on the side of the Si block, while a displacement of 50 nm was imposed at the top of the pillar. Other boundaries were defined as free boundaries. Filet radius was determined as 2.46 $\mu$m shown in Supplementary Figure 15.

\section{Results and Discussion}
\subsection{Rate-dependent compressive properties of copper micropillar up to 500 s$^{-1}$}

An array of pristine copper micropillars was 3D printed using the \mcr AM technique (Figure \ref{fig:01}). The microstructural and chemical analyses on these copper micropillars were conducted using EBSD and EDS, respectively. For the EBSD (Figure \ref{fig:02}) and EDS experiments (Supplementary Figure S2), pillars were lifted out using focused ion beam (FIB)-based milling and placed on a TEM holder (Supplementary Figure S14). Longitudinal cross sections were then cut approximately along the middle of the pillars, and measurements were conducted to characterize the grain boundaries and sizes. Two distinct microstructures were identified in the gage section, namely the microcrystalline (MC, Sample 1, Figure \ref{fig:02}a) and the ultrafine grained (UFG, Sample 2, Figure \ref{fig:02}b) type. The MC pillar is dominated by twin boundaries, while the UFG case mainly has high angle grain boundaries (HAGBs) throughout the gage section (Supplementary Table ST1). Grain size histograms in Figure \ref{fig:02}c reveal that the UFG micropillar has a smaller average grain size ($\sim$170 nm) in the gage section than the MC pillar (with average grain size $\sim$410 nm in diameter). It should be noted that the difference in microstructure (MC and UFG) is due to statistical variations in the neighboring polycrystalline environment during the localized electrodeposition-based 3D printing of micropillars (Supplementary Section SS1: Microstructure variations).

The copper micropillars were then compressed inside an SEM at a wide variety of strain rates from $\sim$0.001 s$^{-1}$ to $\sim$500 s$^{-1}$. The EBSD-based microstructural analysis of the micropillars deformed at 0.1 s$^{-1}$ strain rate are shown in Figure \ref{fig:02}d, e and f. The MC and UFG deformed specimens (Figure \ref{fig:02}d and e), respectively) show similar grain boundary and size characteristics to their non-deformed equivalents. Furthermore, before and after deformation no preferential texture could be identified in the UFG samples (Supplementary Figure S3). 

\begin{figure*}
  \includegraphics[width=\linewidth]{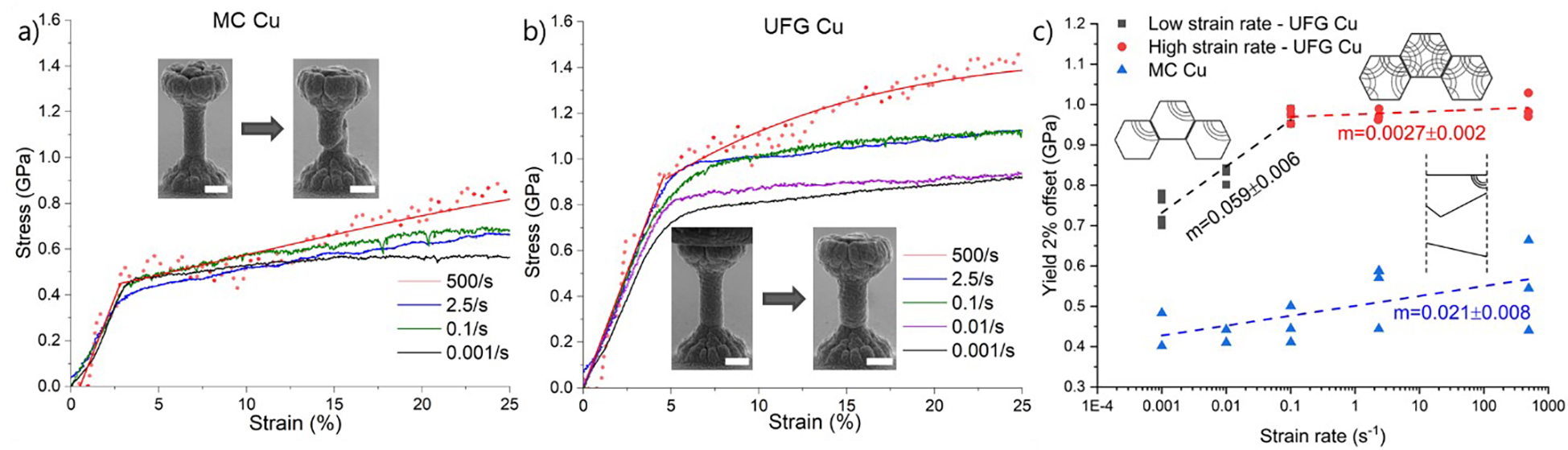}
  \caption{Representative stress-strain curves of copper micropillars with a) MC and b) UFG microstructure with insets of SEM images of copper micropillars before and after deformation (scale bars: 2 \mcr m). c) Extracted yield stress values of copper micropillars as a function of strain rate. The deformation mechanisms postulated for the MC and UFG cases at different strain rate regimes are schematically represented. Insets: Schematic of the hypothesized rate and grain-size dependent deformation mechanisms in copper micropillars.}
  \label{fig:03}
\end{figure*}

The representative stress-strain curves of the copper micropillars compressed at different strain rates are summarized in Figure \ref{fig:03}. As expected, the microstructure, MC or UFG, has a strong influence on the rate-dependent compressive properties of copper micropillars. (Figure \ref{fig:03}a and \ref{fig:03}b).

The MC copper micropillars deform in a single shear-like manner, as seen from the SEM image of the deformed pillar in the inset of Figure \ref{fig:03}a: reminiscent of deformation in single-crystal metal micropillars (Supplementary Video V1 and V3 – MC pillars compressed at strain rates 0.1 s$^{-1}$ and 500 s$^{-1}$) \cite{Imrich.2014}. It is noted that typical micromechanical studies have been conducted on both electrodeposited and FIB-made metal micropillars, with a variety of microstructures including single crystalline \cite{Kiener.2011}, organized twins \cite{Jang.2012} and nanocrystalline \cite{Okamoto.2014}. However, pristine FIB-damage free MC metal micropillars with only a few grains across the diameter, according to our knowledge, have neither been previously fabricated nor mechanically characterized \cite{Kiener.2007, Kiener.2008}. In addition, by conducting rate-dependent micropillar compression tests up to a strain rate of $\sim$500 s$^{-1}$, we observed that the MC micropillars show a weak strain rate dependency (Figure \ref{fig:03}a). Furthermore, during plastic deformation the MC micropillar keeps slipping on the same plane until large surface steps appear and only part of the micropillar top remains attached to the bottom half. In the stress-strain curves shown in Figure \ref{fig:03}a and Supplementary Figure S6, this loss in contact area during plastic deformation manifests itself as softening at large strains above $\sim$20\%, which is especially prominent at lower strain rates. 

On the other extreme, UFG copper micropillars deform in a coherent barreling manner, as representatively shown in the inset of Figure \ref{fig:03}b, at all strain rates (Supplementary Videos V2 and V4 – UFG pillars compressed at strain rates 0.1 s$^{-1}$ and 500 s$^{-1}$). The barreling deformation mode is typical of UFG and nanocrystalline metals and has been reported previously in literature \cite{Zhang.2013}. From the rate-dependent stress-strain signatures of UFG copper from $\sim$0.001 s$^{-1}$ to $\sim$500 s$^{-1}$ (Figure \ref{fig:02}b), we identified that this material shows strong rate-dependency at lower strain rates up to $\sim$0.1 s$^{-1}$ and gets progressively weaker at higher strain rates. In addition, the stress-strain curves exhibit a gradual transition from elastic-to-plastic regime, an indicative sign of inhomogeneous stress distribution in UFG metals due to the microstructural heterogeneity \cite{Rajagopalan.2010}. Similar, grain size dependent deformation behavior in micropillars have been previously identified in copper, tantalum and platinum \cite{Wimmer.2014, Yang.2021, Gu.2012}.

It should be noted that a few samples with intermediate grain size were also identified, resulting in a mixed mode of compressive behavior combining both the characteristics of shear-like and barreling deformation modes (Supplementary Section SS2: Figure S4 and S5, Supplementary Table ST2 and Supplementary Video V5). Due to such ambiguous nature of mixed mode deformation, the current study focuses on pillars that exhibit single-shear like and barreling deformation \emph{i.e.}, with pure MC and UFG microstructure. The differentiation between the three distinct cases of deformation, single-shear like, barreling and mixed mode, is based on the HRSEM inspection of the shape of the deformed pillars.

To confirm the trends of rate-sensitivity quantitatively, the strain rate sensitivity factor $m$ was calculated using the extracted yield strength ($\sigma$) at 2 \% offset strain at different strain rates ($\dot{\varepsilon}$) based on Equation \ref{eq_01} below:

\begin{equation}\label{eq_01}
    m = \frac{\partial \ln \sigma}{\partial \ln \dot{\varepsilon}}
\end{equation}

The yield stress as a function of strain rate is shown in Figure \ref{fig:03}c (Table ST3: comprehensive extracted yield data set). The criterion of 2\% offset strain was chosen, instead of the classical 0.2\% offset strain, in order to systematically avoid errors in yield determination arising from microplasticity in polycrystals due to heterogeneous deformation \cite{Brandstetter.2006}.  The MC copper pillars exhibit a relatively low $m$ value of $0.021 \pm 0.008$ and the rate sensitivity does not change across all studied strain rates. On the other hand, UFG copper micropillars show a higher $m$ value of $0.059 \pm 0.006$ compared to the MC copper micropillars, up to a strain rate of $\sim$0.1 s$^{-1}$. Similarly, a previous macroscopic study by Mao \emph{et al.}\cite{Mao.2018} showed that the strain rate sensitivity of copper increases with decreasing grain size (varied between 90 \mcr m to 500 nm) and this effect is more prominent at strain rates where the viscous drag is minimal (below $\sim$1000 s$^{-1}$. Previously reported strain rate sensitivities for UFG copper based on strain rates up to $\sim$0.1 s$^{-1}$ vary significantly depending on the grain size and the type of grain boundaries from 0.02 to 0.06 \cite{Chen.2006}. The $m$ value obtained in this study for the UFG copper micropillars up to $\sim$0.1 s$^{-1}$ strain rate indeed fall within this range. Traditional strain rate jump tests on the micropillars further confirmed these results (Supplementary Section SS3). At higher strain rates, the UFG copper micropillars exhibit a remarkable yield stress saturation and the strain rate sensitivity decreases by an order of magnitude to $0.0027 \pm 0.002$. Such rate-dependent yield stress saturation has not been observed before in UFG copper micropillars due to the limited range of strain rates ($<$0.1 s$^{-1}$) in previous studies \cite{Malyar.2018}.

\subsection{Atomistic mechanisms responsible for the rate-dependent mechanical properties}

To understand the deformation mechanisms responsible for the mechanical behavior of our copper micropillars, thorough microstructural analysis was carried out on the cross-sections of the deformed pillars (Figure \ref{fig:02}d-f). Regardless of the microstructural type (MC or UFG), there are no statistical differences in the size of grains before and after deformation at a strain rate of $\sim$0.1 s$^{-1}$ (Figure \ref{fig:02}f, Table \ref{table:01} and Supplementary Table ST3). For a larger statistical dataset and to study the same pillar at different cross-sections, multiple FIB slices were prepared $\sim$200 nm apart on a few deformed pillars (Supplementary Table ST4 and Dataset D1 contains the complete statistical summary of the EBSD analysis). 

\begin{table*}[!hb]
\centering
 \caption{Summary of microstructural parameters extracted from the EBSD analysis of the gage section in copper micropillars deformed at 0.1 s$^{-1}$ strain rate. Low angle grain boundaries (LAGB) are defined as: 1$^{\circ}$ -- 10$^{\circ}$, high angle grain boundaries (HAGB) are set as: $>$ 10$^{\circ}$.}\label{table:01}
  \begin{tabular}[htbp]{@{}lllllll@{}}
    \hline
    Deformed & Number of & Min. grain & Max. grain & Special boundary & Relative  & EBSD \\
    sample type & grains & diameter & diameter  & percentages  & frequencies  & mapping\\
    (Name) & (area, [\mcr m$^2$]) &  [\mcr m] & [\mcr m]  & (top two type) & of GBs & step size [\mcr m]\\
    \hline
    MC & 46 & 0.14 & 1.08 & $\Sigma$3: 26.1\% & HAGB: 50.5\%& 0.04 \\
    (Sample 3) & (6.5) &  &  & $\Sigma$9: 1.65\% & LAGB: 49.5\%&  \\
    \hline
    UFG & 116 & 0.14 & 0.42 & $\Sigma$3: 13.2\% & HAGB: 89.9\%& 0.04 \\
    (Sample 4) & (4.4) &  &  & $\Sigma$9: 2.31\% & LAGB: 10.1\%&  \\
    \hline
  \end{tabular}
\end{table*}

An important point to note regarding the copper micropillars is the external pillar dimension-to-average internal grain size ratio ($\eta$). For the MC micropillars, $\eta$ is approximately estimated to vary from 1 to 5 and  for the UFG micropillars $\eta$ is $\sim 10-12$. It has been shown previously that $\eta$ can significantly affect the mechanical properties of micro and nanocrystalline materials.

\subsubsection{MC copper micropillars}

\begin{figure*}
  \includegraphics[width=\linewidth]{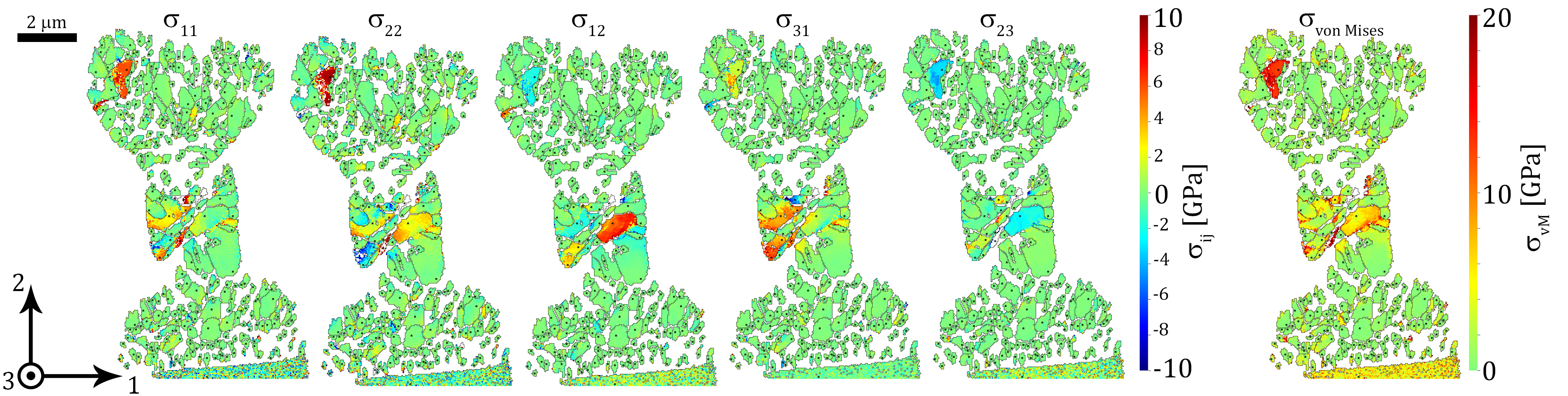}
  \caption{Stress tensor elements by HR-EBSD analysis of the deformed MC copper micropillars, revealing high residual stresses in the sheared large grains.}
  \label{fig:04}
\end{figure*}

HR-EBSD analysis on the deformed MC copper pillar showed that the highest elastic stress concentration in both normal and shear $\sigma$ tensor components was in fact detected in the gage section, close to where the surface shear had occurred (Figure \ref{fig:04}). Hence, the single-shear like deformation was localized to a couple of grains where yielding occurred (also visible in the von Mises stress plot). For the HR-EBSD evaluation, the $\sigma_{33}$ component was assumed to be zero because of the presence of free surface. It is important to note that the stress values presented in Figure \ref{fig:04} are only relative due to the lack of stress-free reference patterns, therefore only relative stress-gradients can be identified using this method. To demonstrate the effect of inappropriate reference allocation, one grain in the upper left corner has been intentionally evaluated using a non-ideal reference, exhibiting high stress values specific to that grain.

From the EBSD data presented (Table \ref{table:01}), the relative frequencies of LAGB and HAGB in the gage section of the deformed pillar are significantly different compared to the undeformed case with similar microcrystalline grain size (Supplementary Table ST1). Increased LAGB presence in the deformed sample (almost 50 \%) is due to the shear-like deformation that created geometrically necessary dislocation (GND) pileups. These GNDs can be detected with the HR-EBSD technique that confirms their increased presence in the deformed pillar with microcrystalline grains (Supplementary Figure S8). 

Apparent activation volume along with the strain rate sensitivity factor are considered as effective kinetic signatures of deformation mechanisms \cite{Krausz.1976}. Thus, we calculated the apparent activation volume ($\Omega$) using Equation \ref{eq:02} at the point of yield, where it is assumed that the microstructure of the MC copper micropillars is similar between tests conducted from strain rates of $\sim$0.001 s$^{-1}$ to $\sim$500 s$^{-1}$. 

\begin{equation}\label{eq:02} 
    \Omega = \sqrt{3} k_B T\frac{\partial \ln \dot{\varepsilon}}{\partial \sigma},
\end{equation}
where $k_B$ is the Boltzmann constant and T is the temperature. Assuming that the Burgers vector $\mathbf{b}$ for a perfect dislocation in copper is 0.255 nm, we calculated an activation volume of $5.55 \pm 2.83$ $b^3$. Such small activation volumes in the range of $1 - 10$ $b^3$ in MC copper micropillars suggest that the deformation is controlled by dislocation nucleation. It is hypothesized that a heterogeneous dislocation nucleation event occurs at yield from the free surface of the copper micropillar, most probably at the coherent twin boundary (CTB) interface of a large grain, which behaves like a single crystal due to the relatively unconstrained boundaries. As mentioned before, given the low $\eta$ of the MC pillars have 1 to 5 grains across the diameter, making them susceptible for free-side surface-based deformation, as the fraction of grains at the surface increases with decreasing $\eta$ \cite{Gu.2012, Zhang.2014}. It has been shown previously via simulations that partial dislocations preferentially nucleate at the intersection of the twin boundary and the face centered cubic (FCC) nanowire surfaces  \cite{Jennings.2011, Deng.2009}. Though other studies suggest that such dislocation nucleation can occur via free surfaces and has been identified FCC metal nanowire simulations \cite{Raj.2016, Weinberger.2008}. For example in penta-twinned, bicrystalline and single-crystalline FCC nanowires it has been shown that the deformation is initiated via a partial dislocation nucleation from the free surfaces or vertices of the nanowire \cite{Filleter.2012, Narayanan.2015, Cheng.2017, Cheng.2020}. Therefore, further micromechanical experiments with very high time and spatial resolution are necessary to experimentally confirm the origin of dislocation nucleation. Though the yield stresses obtained for MC copper micropillars are high ($\sim$0.4 GPa), homogeneous dislocation nucleation could be discounted as a possible deformation mechanism, as even higher stress levels of $4-18$ GPa depending on the crystal orientation are required for such a mechanism to operate in copper \cite{Tschopp.2007}.

In a recent study by Niu \emph{et al.}\cite{Niu.2022}, single crystalline copper nanopillars with 100-800 nm diameter were compressed at quasi-static strain rates of 0.001 s$^{-1}$. The study reported the flow stress at 3\% strain to be between 200-600 MPa, scaling inversely with the pillar diameter. In comparison, in the current study with 2 $\mu$m diameter MC copper micropillars, the yield stress of $\sim$400 MPa was obtained. The higher stress value is primarily due to the Hall-Petch effect, owing to the $\sim$410 nm diameter grains characterized in the current MC pillars. In addition, the FIB damage-free fabrication of the copper pillars presented in the current survey could also enhance the strength of the micropillars. It should also be noted that the reason no apparent stress drops were detected in the stress-strain curves (in Figure \ref{fig:03})) is mainly due to the true-displacement control (enabled due to the intrinsically high mechanical stiffness), offered by the \emph{in situ} micromechanical testing system employed in the present work.

\subsubsection{UFG copper micropillars}

UFG copper micropillars in the current study have an average grain size of $\sim$170 nm and have a high percentage ($\sim$72.5 \%) of HAGBs including $\sim$19.5\% texture-free CTBs. Such UFG micropillars show a high strain rate sensitivity of 0.059 up to a strain rate of $\sim$0.1 s$^{-1}$. Beyond $\sim$0.1 s$^{-1}$, due to the saturation in yield stress, the strain rate sensitivity drops an order of magnitude to 0.0027, signifying a change in deformation mechanism. To understand the rate-controlling mechanisms, the experimental data was fitted to extract the activation volumes in the low strain rate (LSR) regime from $\sim$0.001 s$^{-1}$ to $\sim$0.1 s$^{-1}$, and the high strain rate (HSR) regime from $\sim$0.1 s$^{-1}$ to $\sim$500 s$^{-1}$. The activation volume in the LSR regime was calculated as $7.99 \pm 0.83$ $b^3$, while it increases to $33 \pm 24.8$ $b^3$ at the HSR regime. In the LSR regime, the low activation volume in the range of $1 - 10$ $b^3$ again points to individual dislocation nucleation at yield. It has been shown in literature that the preferred pathway for deformation in low stacking fault energy metals such as UFG and nanocrystalline copper is typically a nucleation of a partial dislocation from triple points \cite{Asaro.2005}. Therefore, it is hypothesized that dislocations nucleate from the HAGBs and weakest triple points in the UFG copper micropillars during deformation. We expect negligible contributions from other deformation processes such grain boundary sliding/diffusive processes such as Coble creep during deformation, as these mechanisms are typically active only at strain rates less than $10^{-4}$ s$^{-1}$ and require a strain rate sensitivity of $\sim$1, which is much higher than what was identified for UFG copper in the current study \cite{Wei.2008a, Wei.2008b}. In addition, the activation volume corresponding to atomic diffusion in the boundary is of the order of $1$ $b^3$, which is still smaller than the measured activation volume for the present UFG copper \cite{Guduru.2007}. Thus, we hypothesize that dislocation-based plasticity is the controlling deformation mechanism for the UFG copper tested here. It is important to note that the FIB-less manufacture of UFG copper micropillars used in this study is critical to ascertain their mechanical behavior unambiguously, as previous studies have shown that both Ga$^+$ and Xe$^+$ ion-based milling of UFG metal micropillars can significantly influence their deformation behavior due to the formation of surface defects such as dislocation loops or surface amorphization \cite{Kiener.2007, Zhang.2014, Xiao.2017}.

In the HSR regime, the activation volume increases by $\sim$4 times compared to the LSR regime, which nominally means a larger number of atoms was involved in the deformation event at yield \cite{McPhie.2012}. In addition, taking into account the high stress levels of $\sim$1 GPa in the HSR regime, we hypothesize that the deformation process at yield possibly transitions to a simultaneous collective dislocation nucleation mechanism. The activation volume of $\sim$ $33.0 \pm 24.8$ $b^3$ calculated at the HSR regime falls in the range of $10 - 100$ $b^3$, which has been previously associated with collective dislocation dynamics process \cite{Jennings.2011}. Khantha \emph{et al.} \cite{Khantha.2001} postulated an alternative dislocation generation theory for crystal deformation, which states that instead of Frank-Read type source, a cooperative nucleation of dislocation can operate at finite temperatures, if the stress levels are high enough. Such a cooperative dislocation mechanism has been previously used to explain the large stress drops at yield during the tensile deformation of pristine microscale copper whiskers \cite{Brenner.1957}. This theory could be adapted to explain the saturation of yield stress at high strain rates ($>$0.1 s$^{-1}$) in our current study of UFG copper (see later  in Figure \ref{fig:05}). At HSR regime when the stress-levels are high enough (above $\sim$0.95 GPa) the collective dislocation nucleation-based deformation mechanism could be favored over the individual dislocation nucleation-based deformation at LSR regime. When such a high stress is reached beyond a strain rate of $\sim$0.1 s$^{-1}$, we posit that multiple triple points within the grain become equally susceptible to dislocation nucleation. Also, in a previous experimental study on single-crystalline and bicrystalline silver nanowires the authors have experimentally shown that silver nanowires undergo a rate-dependent brittle-to-ductile transition at a strain rate of $\sim 0.1 - 273$ s$^{-1}$. This was attributed to a switch in deformation mechanism from a localized failure due to dislocation nucleation from weakest surface site, to a collective dislocation nucleation throughout the nanowire at high strain rates, which resulted in interactions between the dislocations and increased the plastic strain to tensile failure \cite{Raj.2016, Raj.2017}.

\begin{figure*}[!ht]
\centering
  \includegraphics[width=0.99\linewidth]{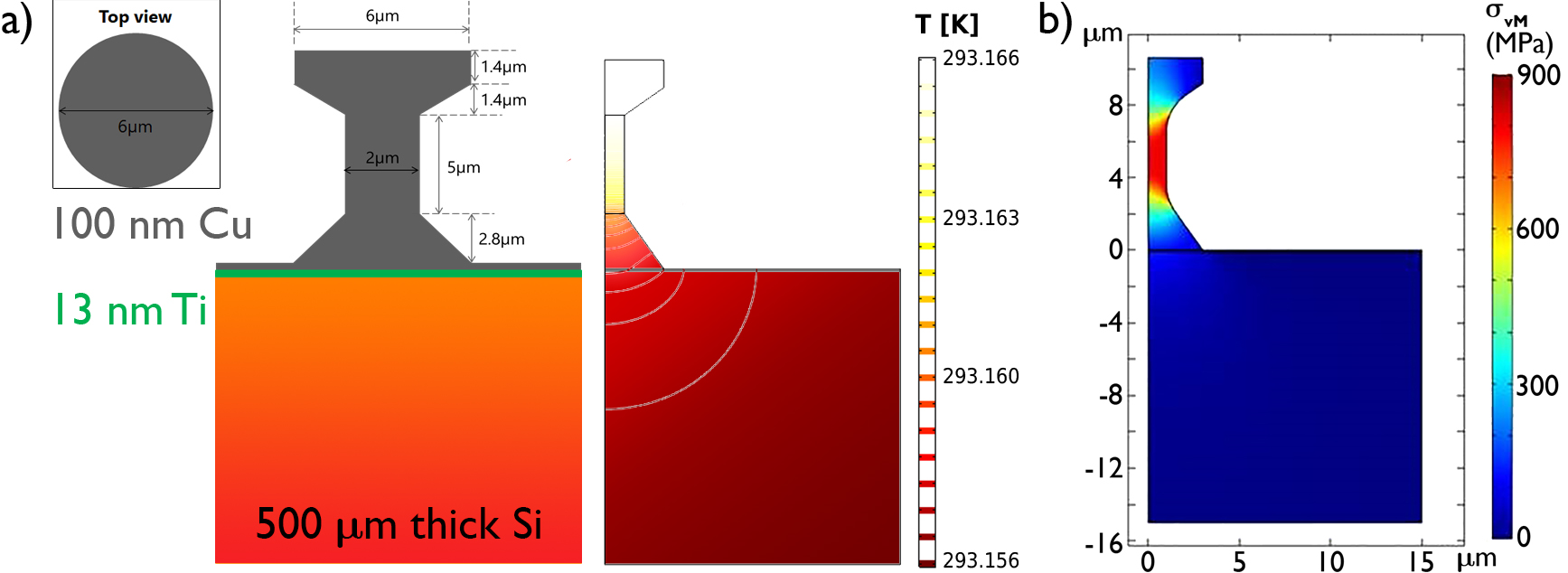}
  \caption{a) FEM-based thermal simulations showing negligible temperature increase in the copper pillars even at high strain rates of 500 s$^{-1}$. b) FEM-based axisymmetric simulations showing homogeneous von Mises stress distribution along the gage section of the specimen.}
  \label{fig:06}
\end{figure*}

Furthermore, the collective nucleation mechanism explains the significant increase in hardening modulus from $\sim$1 GPa to $\sim$2.75 GPa at strain rate of $\sim$500 s$^{-1}$ (Supplementary Figure S9) as evident from the stress-strain signatures in Figure \ref{fig:03}b. Given that the dislocation density is expected to be higher in the HSR regime, the increase in hardening modulus is attributed to a combination of interactions among the simultaneously nucleated multiple dislocations within the relatively large grains of $\sim$170 nm in diameter, and between the dislocations and HAGB/CTBs. It has been previously established that CTBs are better suited for accumulating interfacial dislocations than GBs \cite{Jeon.2015}. In addition, several studies on nanocrystalline metals with a grain size of $\sim 20 - 40$ nm have shown that a smaller grain size prevents pile-up and dislocation tangles, and hence hampers the material to strain harden, even at high strain-rates \cite{Wang.2004}. Specifically, for copper it has been previously established that the threshold grain size for dislocation pile-ups and multiplications to occur within the grain is $\sim$70 nm \cite{Legros.2000}. Thus, the higher hardening modulus of UFG copper micropillars at high strain rates could be attributed to their grain size of $\sim$170 nm and high percentage of CTBs.

A FEM thermal analysis was conducted (Figure \ref{fig:06}a) to understand the extent of adiabatic heating in the micropillars based on the power input during a high strain rate test. Only a negligible temperature increase of $\sim0.01$ K was identified in the micropillar, even when assuming an ideal insulative boundary between the indenter tip and the pillar top. This allowed us to exclude dynamic recrystallization due to adiabatic heating in the UFG copper micropillars within the short testing times of the high strain rate test at $\sim500$ s$^{-1}$. In a similar study, Thevamaran et al. showed that impact testing of single crystal silver nanocubes at ultra-high strain rates of $\sim 10^6$ s$^{-1}$ result in the creation of nanocrystalline grains with high elastic strains \cite{Thevamaran.2016}, which further coarsen over time ($\sim$44 days) to microcrystalline grains due to continuous static recrystallization. In summary, the observations of static recrystallization in UFG copper compressed at high strain rates suggest a new microstructural evolution path from ultrafine grained towards microcrystalline due to the large number of dislocations generated and stored within the grains, and this can serve as a potential method to fabricate copper with gradient nanostructure (Supplementary Section SS5) \cite{Thevamaran.2016}. The validation to unambiguously confirm this phenomenon is out of the scope of this paper, as this requires a systematic microstructure characterization in several samples before and after high strain rate compression.

Figure \ref{fig:06}b plots the simulated von Mises stress distribution near the yield point for the ultrafine grained copper sample. From the modelling it can be concluded that the stress is the highest and homogeneously distributed throughout the gage section, while stress concentration at the rounded corners of the sample is negligible.

\begin{figure*}
\centering
  \includegraphics[width=0.8\linewidth]{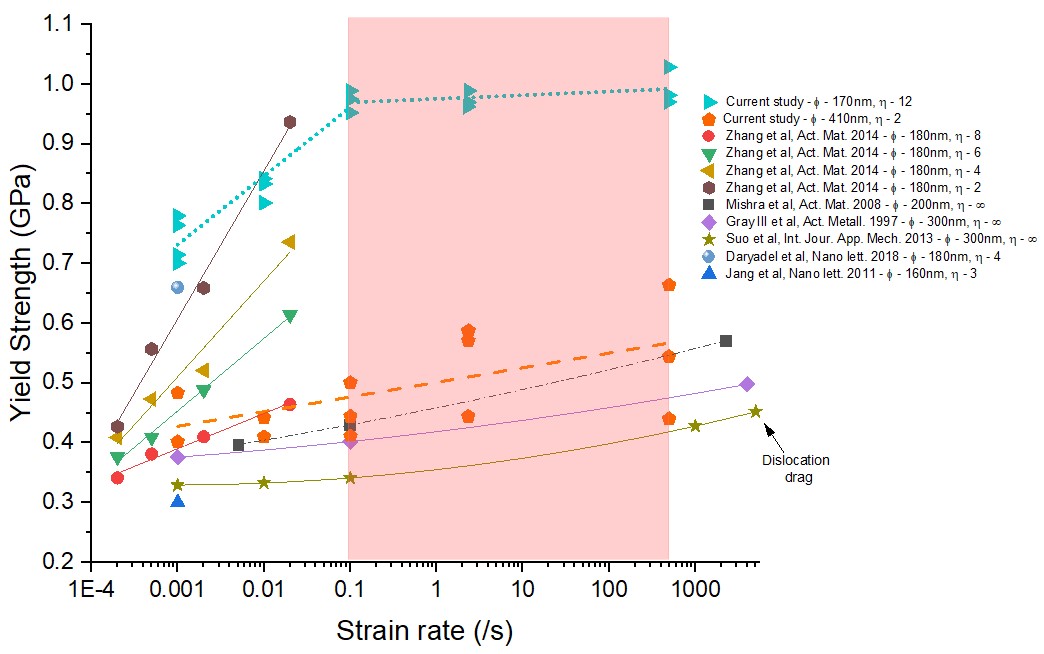}
  \caption{Comparison of rate-dependent results conducted on UFG copper micropillars. Red area highlights the strain rate regime where the current study provides unprecedented experimental results.}
  \label{fig:05}
\end{figure*}

The results obtained in this study is summarized along with previous studies on polycrystalline copper with similar sub-micron sized grains, as shown in Figure \ref{fig:05}. In the case of previous microscale copper tests with UFG microstructure ($\sim$ 180 nm grain diameter), Jhang \emph{et al.} have conducted compression tests upto a strain rate 0.02 s$^{-1}$ on micropillars of varying sizes from 500 nm till 1500 nm. We emphasize here that the results from the currently study at lower strain rates, though with a larger pillar diameter (2000 nm, $\eta = 12$), agree with the results from the 500 nm ($\eta =3$) UFG copper pillars in a previous study, resulting in a high strain rate sensitivity. Previous results on UFG Cu micropillars with a diameter of 1500 nm ($\eta = 8$), show much lower strength. This can be attributed to the pristine nature of the copper micropillars tested in this study, compared to the FIB made micropillars tested by Jhang \emph{et al.} with defective rough walls; a result of the GBs’ weaker resistance to the ion beam compared to the bulk grain interior \cite{Han.2012}. From the comparison of the UFG Cu ($\eta = 12$) and MC Cu ($\eta = 2$) results from the current study it is deduced for micropillars that lower the $\eta$, the lower their strength. Furthermore, macroscale copper tests performed on samples with $\sim$ 200 nm diameter grains exhibit much lower strength at all strain rates. This can be attributed to the heterogeneity and defects in macroscale samples and several orders of magnitude higher $\eta$ ratio ($\eta = \infty$). As such, this also confirms a previously identified cross-over in strength, as the strength increases when the $\eta$ is increased from 2 to 12, while decreases again when the $\eta$ is increased to $\geq 40000$ in macroscale samples. 

In Figure \ref{fig:05}, it is evident that in both microscale and macroscale tests on copper with sub-micron grain size, the strain rates between 0.1 s$^{-1}$  and 1000 s$^{-1}$  remains largely unexplored and the current study fills up this gap by exploring this strain rate regime. As such, the yield stress saturation identified at the strain rates between 0.1 s$^{-1}$  and 500 s$^{-1}$  in UFG Cu has not been ascertained in previous studies. The strength of copper in macroscale tests ($\eta = \infty$) increases significantly beyond a strain rate of $\sim 3000$ s$^{-1}$ due to dislocation drag, while microscale tests at these ultra-high strain rates are further required to confirm such a third deformation regime in samples with lower $\eta$ ratios.

\section{Conclusion}

The microstructure-property relationships of copper micropillars under high strain rates up to 500 s$^{-1}$ were successfully identified using pristine pillars fabricated using metal additive micromanufacturing technique. The stress-strain signatures of the copper micropillars, along with the extracted thermal activation parameters were used to identify the interplay between the grain-size and rate- dependencies on the deformation mechanisms. MC copper micropillars deform via heterogeneous dislocation nucleation from the surface, leading to a single-shear like plastic deformation. UFG copper micropillars deform via single dislocation nucleation, most probably from the HAGB or the triple points in the grain, up to a strain rate of 0.1 s$^{-1}$. At even higher strain rates, a unique rate-insensitivity to yield was identified in UFG copper micropillars. This phenomenon can be attributed to the high-stress and relatively large grain size ($\sim$170 nm) enabled transition to a collection dislocation nucleation-based deformation mechanism. 
In the future, the same µAM technique can be used to fabricate scaled-up full-metal complex 3D microarchitectures including copper microlattices and microsprings. The fundamental dynamic characterization of copper micropillars in the current study will serve as constitutive models to interpret the deformation of such complex mic\-ro\-archi\-tec\-tures.

\section*{Contributions} 
\textbf{R. Ramachandramoorthy:} Conceptualization, Me\-tho\-dology, Validation, Formal analysis, Investigation, Visualization, Writing - Original Draft, \textbf{S. Kalacska:} Conceptualization, Methodology, Software, Validation, Formal analysis, Investigation, Visualization, Writing - Original Draft, \textbf{G. Poras:} Investigation, \textbf{J. Schwiedrzik:} Software, Validation, Visualization, Writing - Original Draft, \textbf{T. Edwards:} Validation, Formal analysis, Investigation, Writing - Original Draft, \textbf{X. Maeder:} Resources, Writing - Review \& Editing, \textbf{T. Merle:} Investigation, \textbf{G. Ercolano:} Investigation, \textbf{W. Koelmans:} Investigation, Writing - Original Draft, \textbf{J. Michler:} Supervision, Writing - Review \& Editing


\section*{Competing interests}  
The authors declare that they have no known competing financial interests or personal relationships that could have appeared to influence the work reported in this paper.

\section*{Acknowledgement}

R.R., S.K. and T.E.J.E. were supported by the EMPA\-POST\-DOCS-II program, which has received funding from the European Union’s Horizon 2020 research and innovation program under the Marie Skłodowska-Curie grant agreement number 754364.

\section*{Supporting Information}

Supporting Information and all data are available in the Main Article and Methods, or from the corresponding authors upon reasonable request.

\bibliography{mybib}

\begin{figure*}[!h]
\textbf{Graphical TOC Entry}\\
\medskip
  \includegraphics[width=0.6\linewidth]{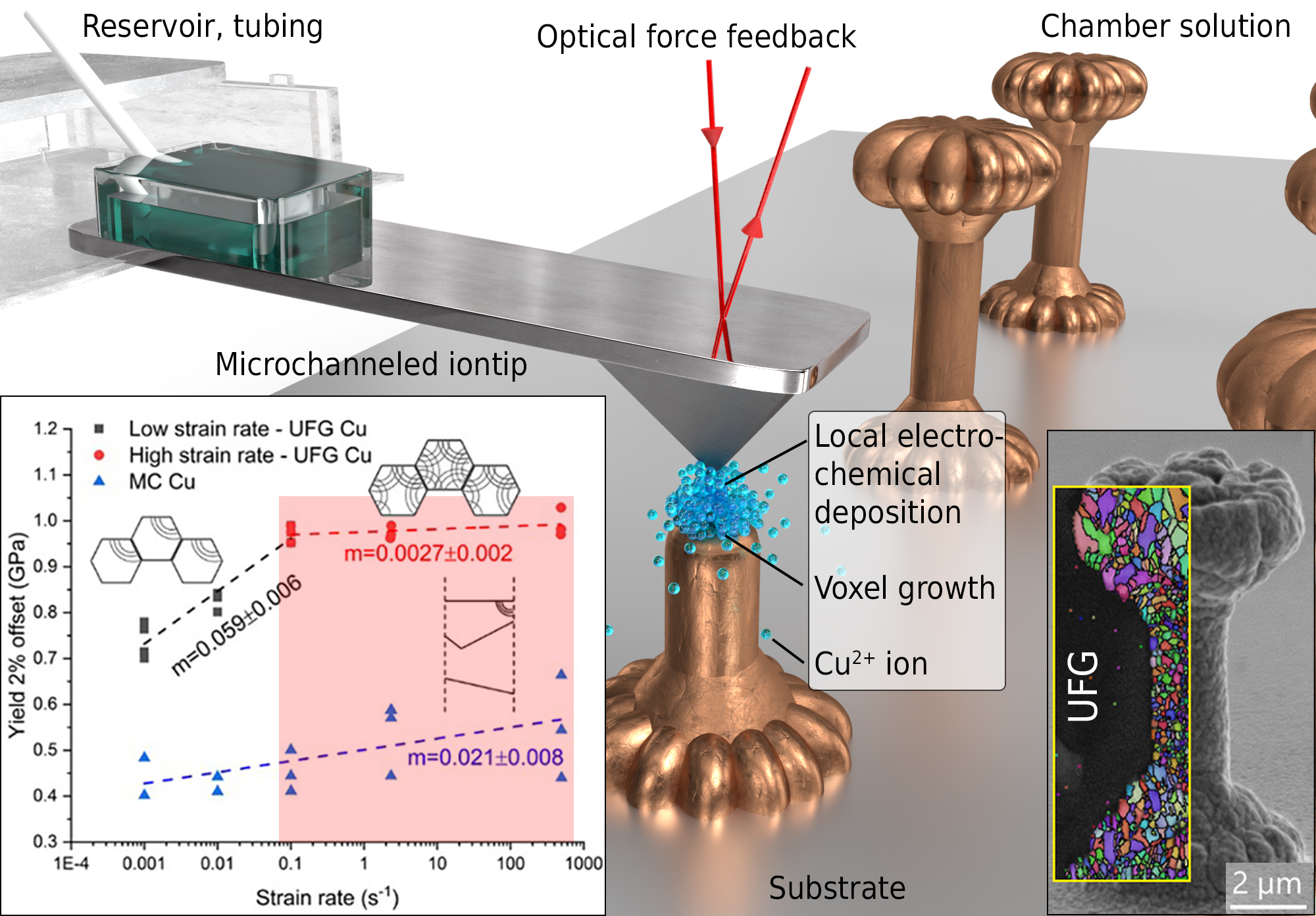}
  \medskip
  \caption*{Dynamic compression testing of pristine 3D printed copper micropillars were performed at various strain rates. Microcrystalline copper pillars deform in a single-shear like manner exhibiting weak strain rate dependency. Ultrafine grained (UFG) pillars deform homogenously via barreling and show a remarkable transition from strong rate-dependence and small activation volumes to yield stress saturation at high strain rates.}
\end{figure*}

\end{document}